\documentclass[11pt,showpacs,showkeys,aps,pra]{revtex4}

\usepackage{amsmath,amssymb}
\usepackage{latexsym}
\usepackage{graphicx}
\usepackage{subfigure}
\usepackage{color}
\usepackage{bm}

\begin{document}

\title{Quantifying Dirac hydrogenic effects via complexity measures}

\author{P.A. Bouvrie$^{a,c}$, S. L\'opez-Rosa$^{b,c}$, J.S. Dehesa$^{a,c}$}
\affiliation{
$^a$Departamento de F\'isica At\'omica, Molecular y Nuclear, Universidad de Granada, 18071-Granada, Spain\\
$^b$Departamento de F\'isica Aplicada II, Universidad de Sevilla, 41012-Sevilla, Spain\\
$^c$Instituto {\em Carlos I} de F\'isica Te\'orica y Computacional, Universidad de Granada, 18071-Granada, Spain}

\email{dehesa@ugr.es}
\date{\today}

\begin{abstract}
The primary dynamical Dirac relativistic effects can only be seen in hydrogenic systems without the complications introduced by electron-electron interactions in many-electron systems. They are known to be the contraction-towards-the-origin of the electronic charge in hydrogenic systems and the nodal disapearance (because of the raising of all the non-relativistic minima) in the electron density of the excited states of these systems. In addition we point out the (largely ignored) gradient reduction of the charge density near and far the nucleus. In this work we quantify these effects by means of single (Fisher information) and composite (Fisher-Shannon complexity and plane, LMC complexity) information-theoretic measures.  While the Fisher information measures the gradient content of the density, the (dimensionless) composite information-theoretic quantities grasp two-fold facets of the electronic distribution: The Fisher-Shannon complexity measures the combined balance of the gradient content and the total 
extent of the electronic charge, and the LMC complexity quantifies the  disequilibrium jointly with the spreading of the density in the configuration space. Opposite to other complexity notions (e.g., computational and algorithmic complexities),  these two quantities describe intrinsic properties of the system because they do not depend on the context but they are functionals of the electron density. Moreover, they are closely related to the intuitive notion of complexity because they are minimum for the two extreme (or least complex) distributions of perfect order and maximum disorder.
\end{abstract}



\maketitle

\section{Introduction}

A major goal of the information theory of the atomic and molecular systems is the quantification of the multiple facets of its internal disorder which is manifest in the electron density of the system, as recently reviewed \cite{sen_11, esquivel_11, esquivel_11b}. First, various single (one-component) information-theoretic measures were used to grasp single facets of the rich variety of complex three-dimensional geometries of the system in a non-relativistic framework, such as the spread of the electronic distribution all over the configuration space (Shannon, R\'enyi and Tsallis entropies), the gradient content (Fisher information) and other manifestations of the non-uniformity of the electron density (disequilibrium). Later, composite (two-component) measures have been proposed to jointly grasp various facets of the electron density. They are called complexity measures because they are minumum for the two extreme distributions of perfect order and maximum disorder (so, approaching the intuitive notion of 
complexity), such as the Cr\'amer-Rao, Fisher-Shannon and LMC (L\'opez-Ruiz, Mancini and Calbet) complexity measures. Opposite to the single-component measures, they are dimensionless (what lets them be mutually compared) and moreover they fulfil a number of invariance properties under replication, translation and rescaling transformations. In addition, contrary to other notions of complexity previously encountered and used in the scientific literature \cite{kolmogorov:pit65, chaitin:jcam66, solomonoff_09, arora_09}, such as the computational and logarithmic complexities which depend on the context, these three complexity measures are intrinsic properties of the system since they are described by density-dependent functionals. Let us also point out that these complexity measures have been bounded from below \cite{lopezrosa:pa09, stam:ic59} and from above \cite{guerrero:pra11}. For further properties of these statistical complexities see the recent monograph of K.D. Sen \cite{sen_11}.

Most of these single and composite information-theoretic quantities have been numerically determined in position space for a great deal of atomic and molecular systems in a Hartree-Fock-like framework (see \cite{sen_11, esquivel_11, esquivel_11b} and references therein). On the contrary, the information theory of relativistic quantum systems is a widely open field \cite{katriel:jcam10, rqi11, peres:rmp04}; indeed, only a few recent works have been done for single-particle systems \cite{manzano:njp10, manzano:epl10, katriel:jcam10} and neutral atoms \cite{borgoo:cpl07, sanudo:pla09, maldonado:pl10, sanudo:irephy09} in various relativistic settings. Let us here mention that the comparison of some Hartree-Fock and Dirac-Fock ground-state results in neutral atoms shows that Shannon entropy is able to characterize the atomic shell structure but it hardly grasps any relativistic effects \cite{borgoo:cpl07}, while the disequilibrium and the LMC complexity measure \cite{borgoo:cpl07}, as well as the Fisher 
information \cite{borgoo_11}, strongly exhibits them. Moreover, it has been recently shown that these quantities are good relativistic indicators for ground-state hydrogenic systems in a Dirac setting \cite{katriel:jcam10} and for ground- and excited states of pionic systems in a Klein-Gordon setting \cite{manzano:njp10, manzano:epl10}.

In this work we use the Fisher information and the Fisher-Shannon and LMC complexity measures to characterize and quantify some fundamental features \cite{grant_07,burke:pps67} of the stationary solutions of the Dirac equation of hydrogenic systems; namely, the well established charge contraction towards the nucleus in both ground and excited states, the raising of all the non-relativistic minima and the (largely ignored) gradient reduction near and far the nucleus of the electron density of any excited state of the system.

The structure of the paper is the following. First, in Section \ref{section2}, we briefly discuss the two information-theoretic quantities needed for this work, and we give the known relativistic (Dirac) and non-relativistic (Schr\"odinger) electron densities of a hydrogenic system, which are factorizable in both frameworks. Second, in Section \ref{section3}, we carry out a detailed study of the dependence of the previous complexity measures on the nuclear charge in the ground state, as well as the quantification of the main dynamical relativistic effects (charge contraction towards the nucleus, minima raising or nodal disappearance, and gradient reduction near and far the nucleus) by means of the Dirac-Schr\"odinger complexity ratios of LMC and Fisher-Shannon types. Third, in Section \ref{section4}, we analyse the dependence of the complexity measures on the energy and the relativistic quantum number as well as the associated information planes for the ground and various excited states. Finally, some 
conclusions are given.

\section{Complexity measures and Dirac hydrogenic densities: Basics}
\label{section2}

In this Section we briefly discuss the concepts of LMC shape complexity \cite{lopezruiz:pla95,anteneodo:pla96} and Fisher-Shannon complexity \cite{angulo:pla08,romera:jcp04} of a general probability density $\rho(\mathbf{r})$ used in this paper, which turn to be good indicators of the Dirac relativistic effects in hydrogenic systems. Then, we collect here the known Dirac wave funtions of the hydrogenic bound states, and their associated probability densities together with its non-relativistic limit (Schr\"odinger densities).

The LMC shape complexity \cite{lopezruiz:pla95, anteneodo:pla96} is defined by the product of the so-called disequilibrium $D \left[ \rho \right]$ (which quantifies the departure of the probability density from uniformity) and the exponential of the Shannon entropy $S\left[ \rho \right]$ (a general measure of the uncertainty of the density):
\begin{equation}\label{eq:def_LMC}
C_{\text{LMC}} \left[ \rho \right] = D\left[ \rho \right] \times e^{S\left[ \rho \right]},
\end{equation}
where
\begin{equation}\label{eq:def_D-S}
D\left[ \rho \right] = \int \left[ \rho(\mathbf{r}) \right]^2 d\mathbf{r}; \quad S\left[ \rho \right] =  -\int \rho(\mathbf{r}) \ln \rho(\mathbf{r}) d\mathbf{r}.
\end{equation}

The Fisher-Shannon complexity \cite{angulo:pla08,romera:jcp04} is given by 
\begin{equation}\label{eq:def_FS}
C_{\text{FS}} \left[ \rho \right] = I\left[ \rho \right] \times J \left[ \rho \right],
\end{equation}
where 
\begin{equation}\label{eq:def_I-J}
I\left[ \rho \right] = \int \frac{|\bar\nabla\rho (\mathbf{r})|^2}{\rho (\mathbf{r})} d\mathbf{r} ; \quad J \left[ \rho \right]= \frac{1}{2 \pi e} e^{\frac{2}{3}S\left[ \rho \right]}
\end{equation}
are the (translationally invariant) Fisher information \cite{frieden_04} and the Shannon entropic power \cite{cover_91} of the probability density, respectively. The latter quantity, which is an exponential function of the Shannon entropy, measures the total extent to which the single-particle distribution is in fact concentrated \cite{cover_91}. The Fisher information, $I\left[ \rho \right]$, which is closely related to the kinetic energy \cite{hamilton:jcam10}, is a local information-theoretic quantity; i.e., it is very sensitive to strong changes on the distribution over a small-sized region of its domain.

On the other hand, the Dirac wavefunctions of the stationary states of a hydrogenic system with nuclear charge $Z$ are described by the eigensolutions $(E,\psi^D)$ of the Dirac equation of an electron moving in the Coulomb potential $V(r)= -\frac{Z e^2}{4\pi\epsilon_0r}$, namely
\begin{equation}\label{eq:dirac_eq}
\left( E + i \hbar c \bm{\alpha} \cdot \bm{\nabla} - \beta m_0c^2 - V(r)\right) \psi^D = 0,
\end{equation}
where $\bm{\alpha} \equiv \left(\alpha_1, \alpha_2, \alpha_3 \right)$ and $\beta$ denote the $4\times4$ Dirac matrices and $m_0$ denotes the rest mass of electron.

The stationary eigensolutions are most naturally obtained by working in spherical polar coordinates and taking into account that the Dirac hamiltonian conmute with the operators $\{{\mathbf{J}}^2, J_z, \mathbf{K}\}$, where the total angular momentum operator $\bf J = \bf L + \bf S$ and the Dirac operator $\mathbf{K} = \beta (\mathbf{\Sigma} \cdot \mathbf{L} + \hbar)$, being $\bf L$ and $\mathbf{S} \equiv \frac{\hbar}{2} \mathbf{\Sigma}$ the orbital and spin angular momenta, respectively. So, the stationary states are to be characterized by the quantum numbers $(n,k,m_j)$, where $n \in \mathbb{N}$, the Dirac or relativistic quantum number $k=\pm 1,\pm 2, ... , -n$ and $-j \leq m_j \leq j$ with $j= \frac{1}{2},\frac{3}{2},...,n-\frac{1}{2}$. Besides, let us remark that $k=\mp(j+\frac{1}{2})$ for $j=l\pm\frac{1}{2}$, so that $k=-(l+1)$ if $j=l+\frac{1}{2}$ and $k=l$ if $j=l-\frac{1}{2}$; in other terms, $k=\pm(j+\frac{1}{2})$, to which there corresponds (upper component) angular momentum $l=j\pm \frac{1}{2}$ 
and (lower component) $l'=j=\mp\frac{1}{2}$. The energy eigenvalues are known (see e.g. \cite{grant_07, greiner_00, drake_96, swainson:jpa95}) to be
\begin{equation}
E = M\left(1+\frac{\left(\alpha Z\right)^2}{\left(n-|k|+\sqrt{k^2-\left(\alpha Z\right)^2}\right)^2}\right)^{-1/2}, 
\end{equation}

where $\alpha$ denotes the fine structure constant, $M = m_0 c^2$ and $Z<137$. For $Z>137$ the Klein paradox \cite{klein:zp29} comes into play and the eigenenergies become complex beyond that point; the resolution of this paradox is known to be related with the creation of electron-positron pairs from de Dirac-Fermi sea \cite{calogeracos:cp99}. Note that, because of the smallness of the binding energies, $E$ is only slightly less than $m_0 c^2$. The corresponding eigensolutions of the bound relativistic hydrogenic states are given by the four-component spinors
\begin{equation}\label{eq:dira_fo}
\psi_{nkm_j}^D(\bm{r}) = \left( \begin{array}{c} g_{nk}(r)\Omega_{km_j}(\theta,\phi)\\ i f_{nk}(r)\Omega_{-k m_j}(\theta,\phi) \end{array}\right),
\end{equation}
where the symbol $\Omega_{k,m_j}(\theta,\phi)$ denotes the (two-component) spin-orbital harmonics
\begin{equation} \label{eq:sphericalspinor}
 \Omega_{km_j}=\left(\begin{array}{c} -\frac{k}{|k|} \sqrt{\frac{k+\frac{1}{2}-m_j}{2 k + 1}} Y_{|k+\frac{1}{2}|-\frac{1}{2},m_j-\frac{1}{2}}(\theta,\phi)\\ \sqrt{\frac{k+\frac{1}{2}+m_j}{2 k + 1}} Y_{|k+\frac{1}{2}|-\frac{1}{2},m_j+\frac{1}{2}}(\theta,\phi)\end{array}\right),
\end{equation}
and the so-called large ($g$) and small ($f$) radial components with the normalization $\int_0^\infty(g^2+f^2) r^2 dr = 1$, are known to be
\begin{align}\label{eq:def_fg}
\hspace*{-2cm} \left. \begin{array}{c} g_{nk}(r)\\ f_{nk}(r) \end{array} \right\}&  =
\frac{\pm(2\lambda)^{3/2}}{\Gamma(2\gamma+1)}\sqrt{\frac{(M\pm E)\Gamma(2\gamma+n'+1)}{4M\frac{(n'+\gamma)M}{E}\left(\frac{(n'+\gamma)M}{E}-k\right)n'!}} (2\lambda r)^{\gamma-1}e^{-\lambda r}   \nonumber \\
& \hspace*{-1.5cm}\times \left[\left(\frac{(n'+\gamma)M}{E}-k\right)F(-n',2\gamma+1;2\lambda r) s  \mp n'F(1-n',2\gamma+1;2\lambda r)\right]
\end{align}
where $n'=n-|k|$, $\gamma=\sqrt{k^2-(\alpha Z)^2}$, $\lambda = \frac{1}{\hbar c} (M^2-E^2)^{1/2}$, and $F(a,b;z)$ denotes the Kummer confluent hypergeometric function. Notice that the lower components of the Dirac wavefunction have an opposite parity than the upper ones. Moreover, the binding energy $B=|E^D_{n,|k|}|=m_0c^2-E$ depends on the principal quantum number $n$ and on the absolute value of the Dirac quantum number $k$, but not on its sign. This means that states with the same angular momentum quantum number $j$ which belongs to different pairs of orbital quantum numbers ($l$, $l'$) are degenerated in energy. In addition we should point out that we will often identify $\psi^D_{nljm_j}$ with $\psi^D_{nkm_j}$ although the Dirac relativistic states are no longer eigenfunctions of the orbital angular momentum because the Dirac hamiltonian does not conmute with $\bf L$; so, the orbital quantum number is not a good quantum number. Indeed, each relativistic state contains two values: $l$ and $l'=l\pm1$. 
However, since the component with the radial function $g_{nk}(r)$ is large as compared to its partner $f_{nk}(r)$, the value $l$ pertaining to the large component may be used to denote the state. Then, although we use the non-relativistic notation $|nljm_j\rangle$ we should keep in mind that it stands for $|nkm_j\rangle$

Then, the Dirac probability density $\rho^D_{nljm_j}(\bm{r}) = | \psi_{nljm_l}^D (\bm{r}) |^2$ of the hydrogenic state $\left| n  l j m_j \right\rangle$ can be written down in the following separable form:
\begin{equation} \label{eq:dirac_dens}
\rho^D_{nljm_j}(\bm{r}) = \rho_{\text{radial}}^D(r) \rho_{\text{angular}}^{}(\theta), 
\end{equation}
where the radial and angular parts are given by
\begin{equation}
\rho_{\text{radial}}^D(r)=|g_{nk}(r)|^2+|f_{nk}(r)|^2 
\end{equation}
and 
\begin{align}\label{eq:spher_dens}
\rho_{\text{angular}}^{}(\theta)& =\langle l,m_j-\frac{1}{2};\frac{1}{2},+\frac{1}{2}|j,m_j \rangle^2 |Y_{l,m_j-\frac{1}{2}}|^2 \nonumber \\
& \quad +  \langle l,m_j+\frac{1}{2};\frac{1}{2},-\frac{1}{2}|j,m_j \rangle^2 |Y_{l,m_j+\frac{1}{2}}|^2, 
\end{align}
respectively.

Finally, it is well known that in the non-relativistic limit of the hydrogenic system the large component $g_{nk}(r)$ tends to the corresponding radial function of the Schr\"odinger equation while the small component $f_{nk}(r)$ tends to cero. So, the Schr\"odinger probability density $\rho^S_{nljm_j}(\bf r)$ which describes the state $\left| n  l j m_j \right\rangle$ of the system is
\begin{equation}\label{eq:Sch_dens}
\rho^S_{nljm_j}(r,\theta)=|\psi^S_{nljm_j}(r,\theta,\phi)|^2=\rho_{\text{radial}}^S(r)\rho_{\text{angular}}^{}(\theta),
\end{equation}
where 
\begin{align}\label{eq:radial_Sch}
\rho_{\text{radial}}^S(r) & = \frac{\Gamma(n-l)}{2n\Gamma(n+l+1)}\left(\frac{2Z}{a_0 n}  \right)^{2l+3} \nonumber \\
\quad & \times e^{-\frac{2Z}{a_0 n}r} r^{2l} \left| L_{n-l-1}^{2l+1}\left( \frac{2Z}{a_0 n} r\right)\right|^2,
\end{align}
gives the radial part of the wavefunction, and $\rho_{\text{angular}}^{}(\theta)$ is the same angular part as in Dirac case given by Eq. (\ref{eq:spher_dens}). The corresponding energy of the non-relativistic system is known to be $E^S_n = - \frac{\hbar^2 Z^2}{2 a_0^2 n^2}$.

\section{Complexity quantification of Dirac effects}
\label{section3}

In this section we quantify the two main dynamical Dirac relativistic effects (charge contraction towards the origin and raising of all non-relativistic minima), as well as the gradient reduction near and far the origin, in hydrogenic systems by means of the LMC and Fisher-Shannon complexity measures. This is done by studying the comparison between the Schr\"odinger and Dirac values of the LMC and Fisher-Shannon complexities of ground and excited states of hydrogenic systems. Specifically we show the dependence of these quantities, as well as the Fisher-Shannon information plane, on the nuclear charge $Z$ and the principal quantum number $n$. For simplicity and convenience we will use atomic units hereafter.

\subsection{Dependence on the nuclear charge}
\label{section3A}

\begin{figure}[ht]
\centering
\subfigure{
\includegraphics[scale=0.625]{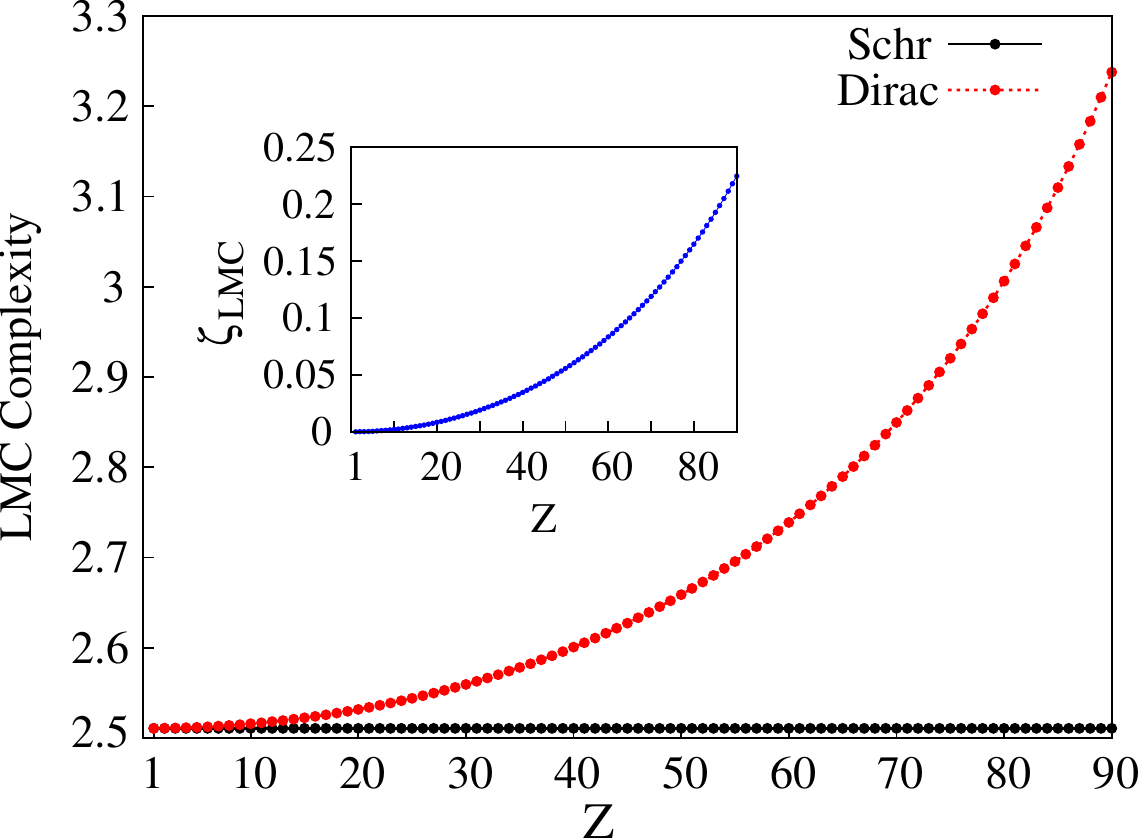} 
\label{fig:lmc_depZ}
}
\subfigure{
\includegraphics[scale=0.625]{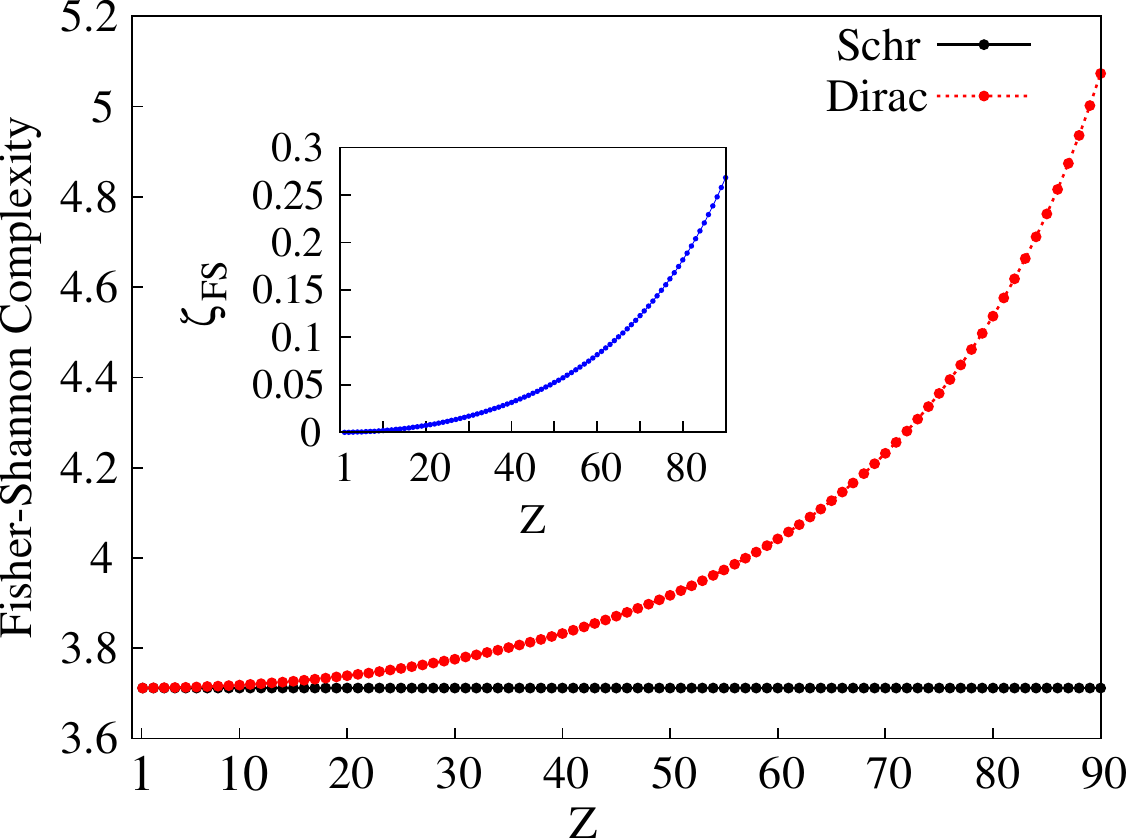} 
\label{fig:fs_depZ}
}
\caption{(Color online) Dependence of the ground-state hydrogenic LMC (left) and Fisher-Shannon (right) complexity measures on the nuclear charge Z}
\label{fig:depZ}
\end{figure}

First, let us present and discuss the dependence on the nuclear charge $Z$ of the LMC (see Fig. \ref{fig:depZ}-left) and Fisher-Shannon (see Fig. \ref{fig:depZ}-right) complexity measures in the ground state of the hydrogenic system in the Schr\"odinger and Dirac settings described in the previous section. We find that for both complexity measures (i) the Schr\"odinger values remain constant for all Z’s (as recently proved in an analytical way \cite{lopezrosa:pa09, dehesa:epjd09}), and (ii) the Dirac values enhance when the nuclear charge is increasing, in accordance with the corresponding Klein-Gordon results found in pionic systems \cite{manzano:epl10}. 

\begin{figure}[ht]
\centering
\subfigure{
\includegraphics[scale=0.625]{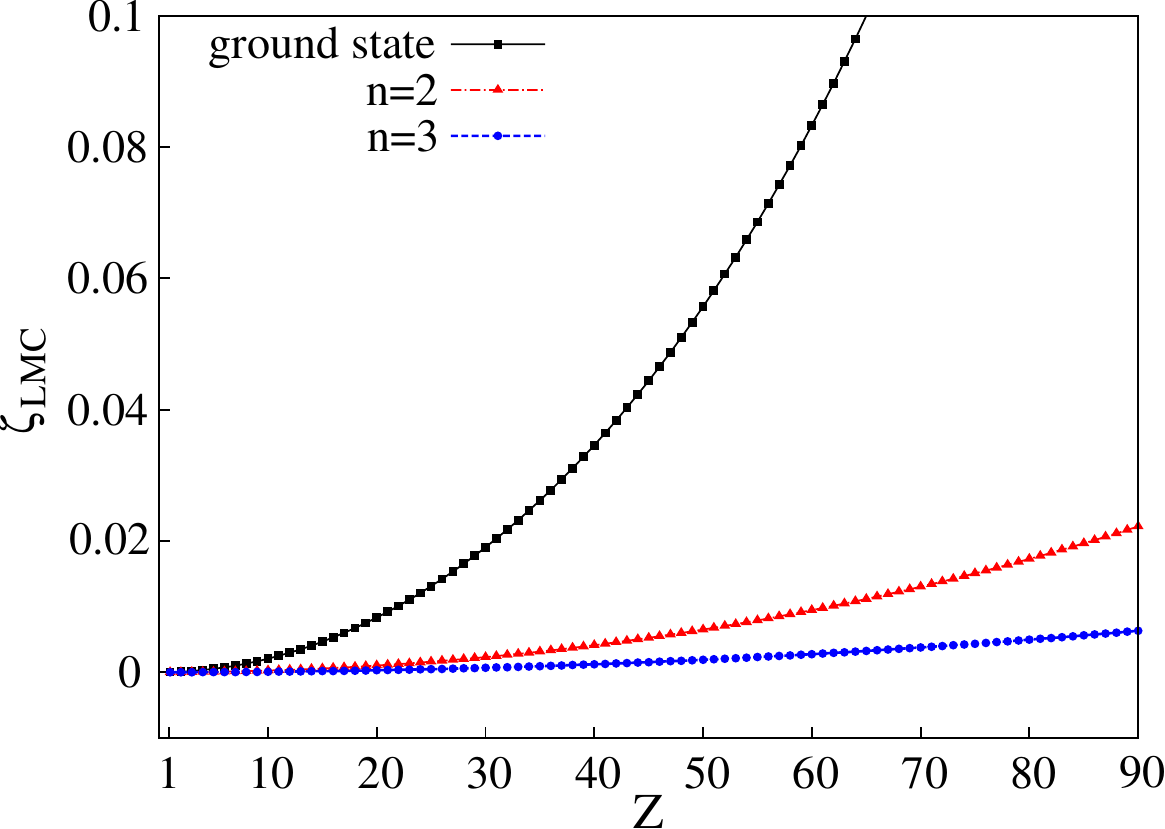}
\label{fig:ratio_lmc_depZ}
}
\subfigure{
\includegraphics[scale=0.625]{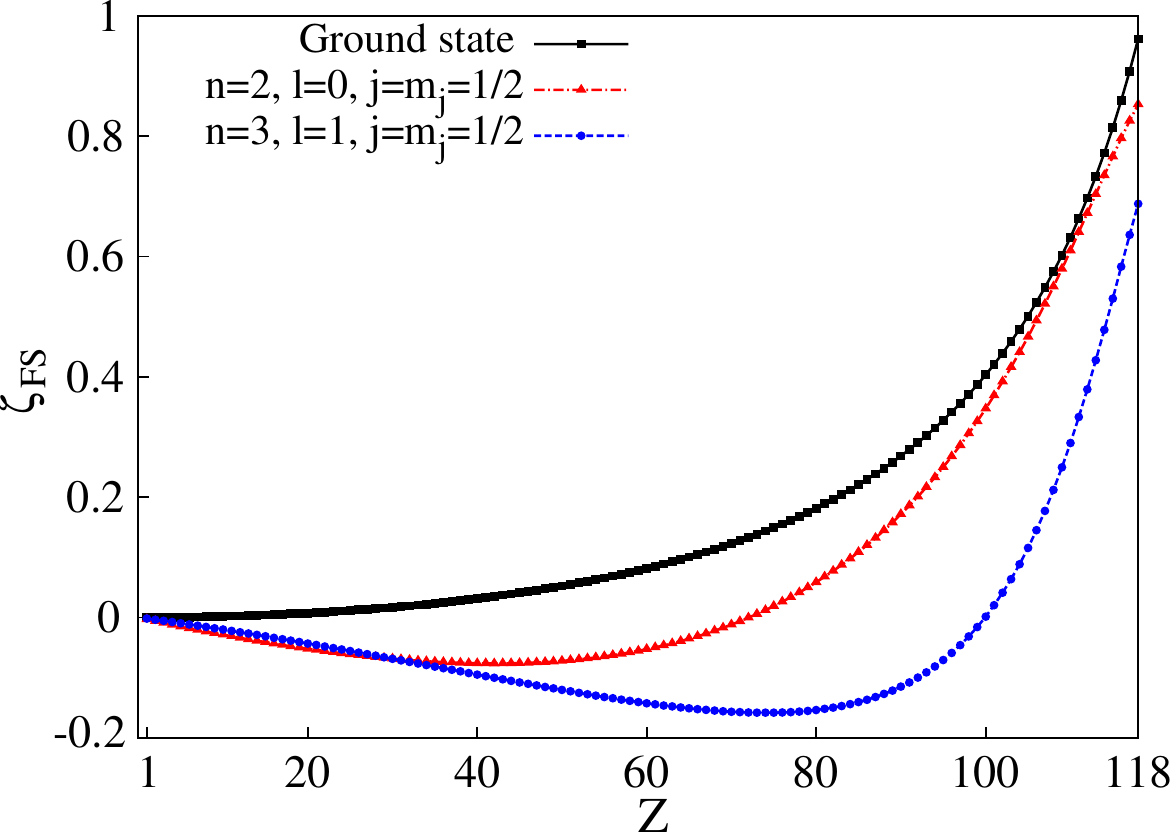} 
\label{fig:ratio_fs_depZ}
}
\caption{(Color online) LMC relative ratio, $\zeta_{LMC}$, for circular states with $n\le 3$ (left) and Fisher-Shannon relative ratio, $\zeta_{FS}$, for the ground state and the excited states $(n,l,j,m_j)=(2,0,\frac{1}{2},\frac{1}{2})$ and $(3,1,\frac{1}{2},\frac{1}{2})$ (right).}
\label{fig:ratio_depZ}
\end{figure}

This enhancement is provoked by the contraction of the electron density towards the origin, a phenomenon similar to that observed for Klein-Gordon single-particle systems \cite{manzano:njp10, manzano:epl10}. To quantify it we have defined the relative ratios $\zeta_{\text{LMC}}=1-\frac{C^{\text{S}}_{\text{LMC}}}{C^{\text{D}}_{\text{LMC}}}$ and $ \zeta_{\text{FS}}=1-\frac{C^{\text{S}}_{\text{FS}}}{C^{\text{D}}_{\text{FS}}}$. They are shown in the inner windows of Figs. \ref{fig:depZ}-left-and-right in terms of $Z$. We observe that both complexity ratios behave similarly in the ground state. This is not, however, the case for other states as it is illustrated in Fig. \ref{fig:ratio_depZ}-left for the LMC measure in three circular states  with $n \leq 3$, and in Fig. \ref{fig:ratio_depZ}-right for the Fisher-Shannon complexity in the ground state and two excited states. Therein we observe that while the LMC ratio is always positive and has an increasing behavior as a function of Z, this is not always the case 
for the Fisher-Shannon ratio. Indeed, notice that the latter ratio can reach negative values for the excited states, indicating that the Dirac value of the Fisher-Shannon complexity is lower than the Schr\"odinger one. 

\begin{figure}[ht]
\centering
\subfigure{
\includegraphics[scale=0.625]{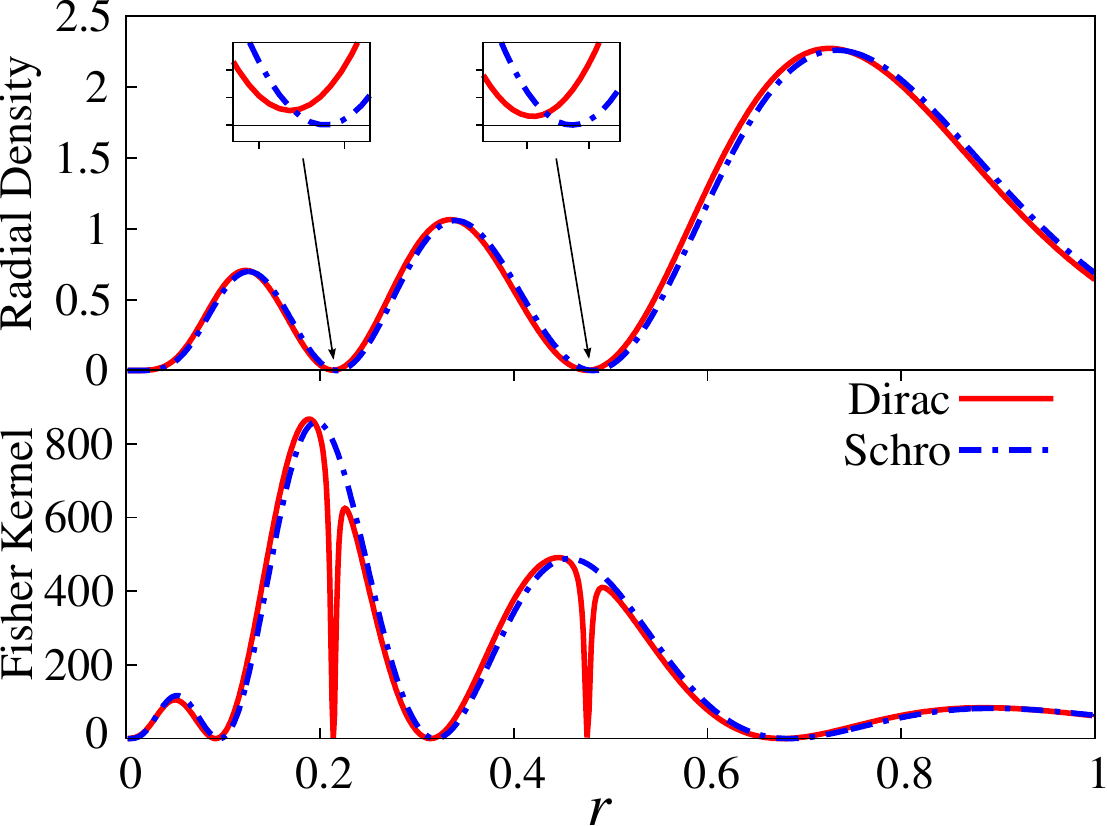}
\label{fig:state_n5l2}
}
\subfigure{
\includegraphics[scale=0.625]{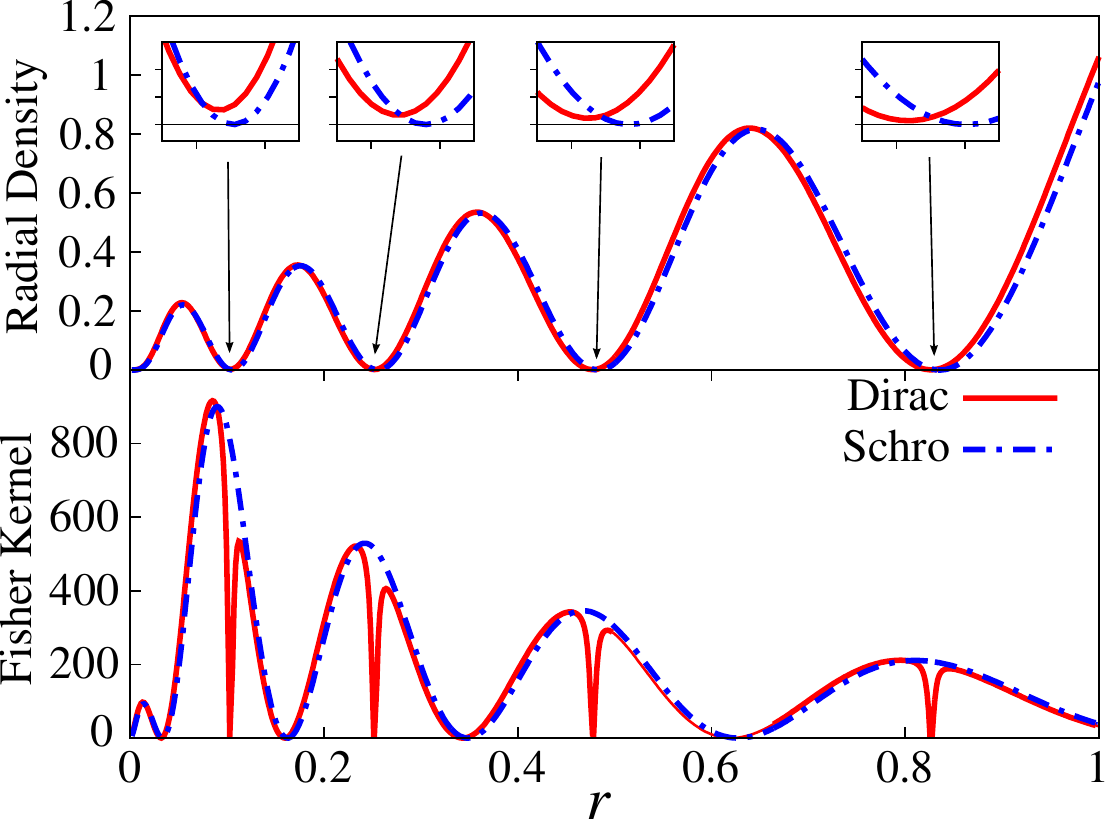}
\label{fig:state_n6l1}
}
\caption{(Color online) Radial density, $D^i(r)$, and radial Fisher information kernel, $I^i_{kernel}(r)$, in the Dirac ($i=D$) and Schr\"odinger ($i=S$) settings for the hydrogenic states $n=5,l=2,j=\frac{1}{2}$ (left) and $n=6,l=1,j=\frac{1}{2}$ (right) with nuclear charge $Z=50$. Atomic units have been used.}
\label{fig:dens_Ikernel}
\end{figure}

The positivity of the LMC ratio can be understood because it measures the charge contraction towards the nucleus by means of two global concentration information-theoretic quantities, the disequilibrium and the Shannon entropy. We observe that although these two factors work in the same sense, the contribution of the disequilibrium turns out to be much greater than that of the Shannon entropy. The negative behavior of the Fisher-Shannon ratio is more difficult to explain, because it quantifies the combined balance of the spreading (via the Shannon entropy) and the gradient content (via the Fisher information) of the charge distribution of the hydrogenic system. To understand this phenomenon we analyze the behavior of the two components of the Fisher-Shannon complexity \eqref{eq:def_FS}. Keeping in mind that the Shannon entropy is not very sensitive to the relativistic effects, the former analysis boils down to a careful determination of the Fisher information which can be written down as
\begin{equation}\label{eq:Fisher_separable}
I\left[ \rho \right]= I_{\text{radial}}+ \left\langle r^{-2} \right\rangle \times  I_{\text{angular}}, 
\end{equation}
where $I_{\text{radial}}$ denotes the Fisher information of the radial probability function $\rho_{\text{radial}}^i(r)$ ($i=D$ or $S$ in the Dirac and Schr\"odinger case, respectively), and $I_{\text{angular}}$ the Fisher quantity associated to the angular probability function, $\rho_{\text{angular}}^{}(\theta)$. Let us point out that the Fisher information presents a singularity at $Z=118.68$, as pointed out by Katriel \& Sen \cite{katriel:jcam10}, what explains why we do not go to the extreme relativistic limit. Since the angular density is the same function in both relativistic and non-relativistic descriptions and $\left\langle r^{-2} \right\rangle$ is slightly higher in the relativistic case, the main reason for the negativity of the Fisher-Shannon complexity ratio arises from the difference between the Dirac and Schr\"odinger radial probability densities. This is clearly shown in Fig. \ref{fig:dens_Ikernel}, where we have plotted the Dirac and Schr\"odinger radial densities, $D^i(r)\equiv\rho_{\text{
radial}}^{i}(r) r^2$, and the corresponding Fisher kernel, $I^i_{kernel}(r)\equiv\frac{1}{\rho_{\text{radial}}^{i}(r)}\left ( \frac{\partial \rho_{\text{radial}}^{i}(r)}{\partial r}\right )^2 r^2$, for the excited states $(n, l, m_j) = (5, 2, 3/2)$ and $(6,1, 3/2)$ of the hydrogenic atom with nuclear charge $Z = 50$ in the left and right sides, respectively.
 
Therein we notice that while the Schr\"odinger radial density, $D^S(r)$, vanishes at various points (nodes), the Dirac radial density, $D^D(r)$, only vanishes at the origin and the infinity. This means that the Dirac density have finite values also at the radial positions of the non-relativistic nodes (this is the relativistic minima-raising effect). Hence, the Fisher information kernel is zero at these points because although the radial density does not vanish, its derivative does (see Fig. \ref{fig:dens_Ikernel}); and this is the reason for the high negative values of $\zeta_{FS}$ detected in Fig. \ref{fig:ratio_depZ}.

\begin{figure}[ht]
\centering
\subfigure{
\includegraphics[scale=0.625]{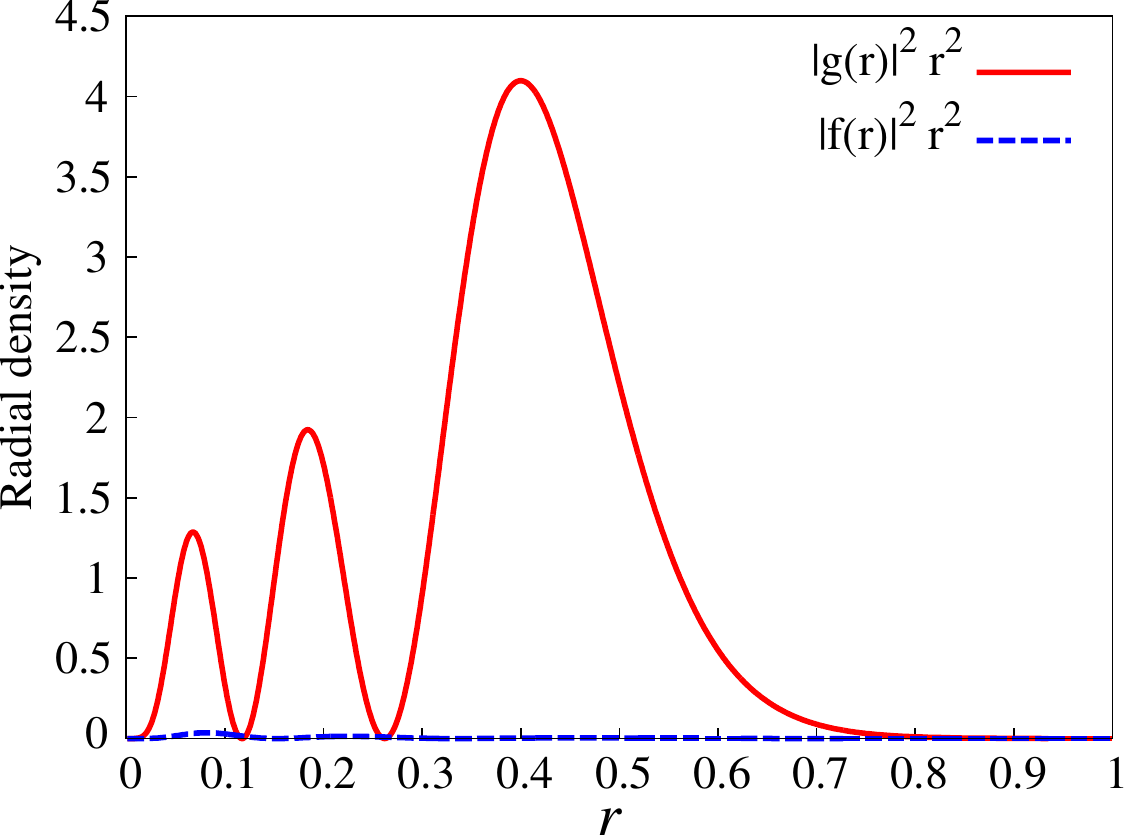}
\label{fig:dens_Z90_fg}
}
\subfigure{
\includegraphics[scale=0.625]{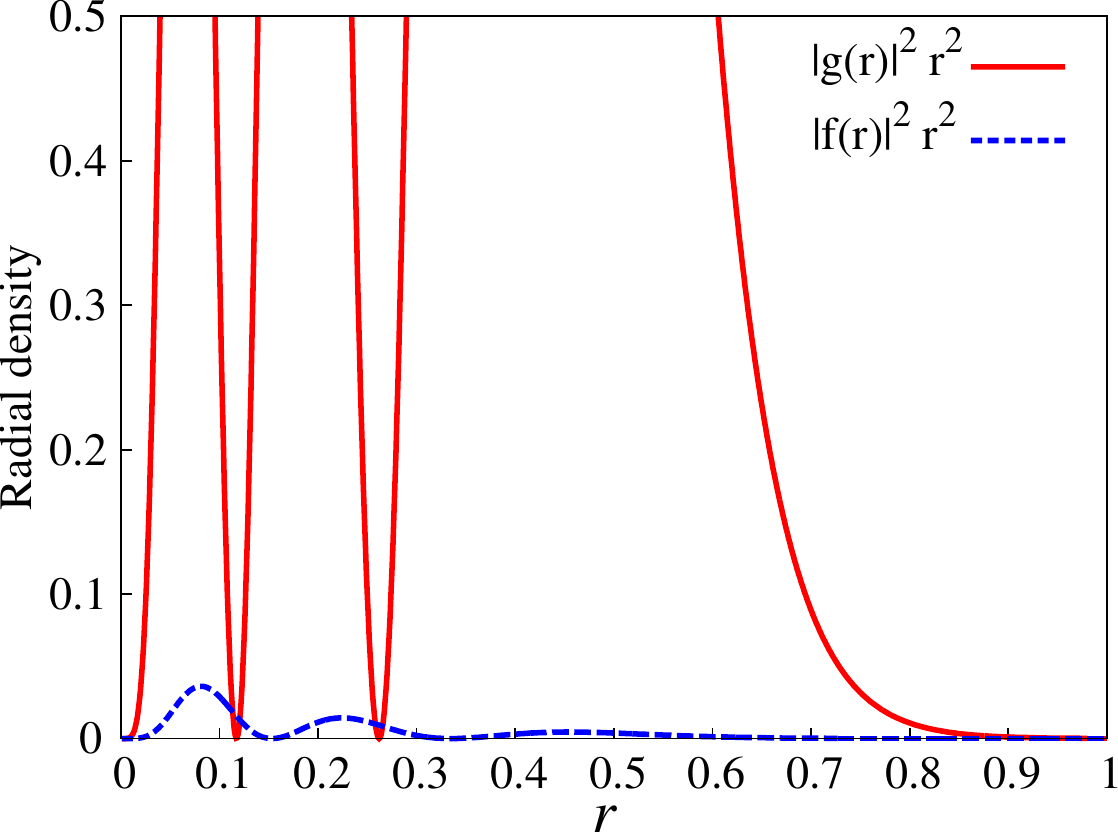}
\label{fig:dens_Z90_fg_zoom}
}
\caption{(Color online) Contribution of the $g(r)$ and $f(r)$ to the total probability density for the hydrogenic state $n=5,l=2,j=2.5$ with nuclear charge $Z=90$. Atomic units have been used.}
\label{fig:dens_Z90}
\end{figure}

This relativistic effect of nodal disapearance (or existence of non-nodal minima) in the Dirac density, firstly pointed out by  Burke and Grant \cite{grant_07,burke:pps67}, is indeed due to the different behavior of the two components $g(r)$ and $f(r)$ of the Dirac wave function. Both functions vanish at different values of $r$ as we can observe in Fig. {\ref{fig:dens_Z90}}, where the contribution of $g$ and $f$ to the total probability density for the state $n=5,l=2,j=2.5$ of the system with $Z=90$ has been plotted for illustrative purposes. As we can see in Fig. \ref{fig:dens_Z90}, the largest contribution to the total probability density is indeed due to the component $g(r)$ of the Dirac spinor \eqref{eq:dira_fo}. The contribution of the $f$-component, although very small, is sufficiently significant as to make the Dirac density not to vanish for all radial values except for $r = 0$ and $\infty$.

\begin{figure}[ht]
\centering
\subfigure{
\includegraphics[scale=0.625]{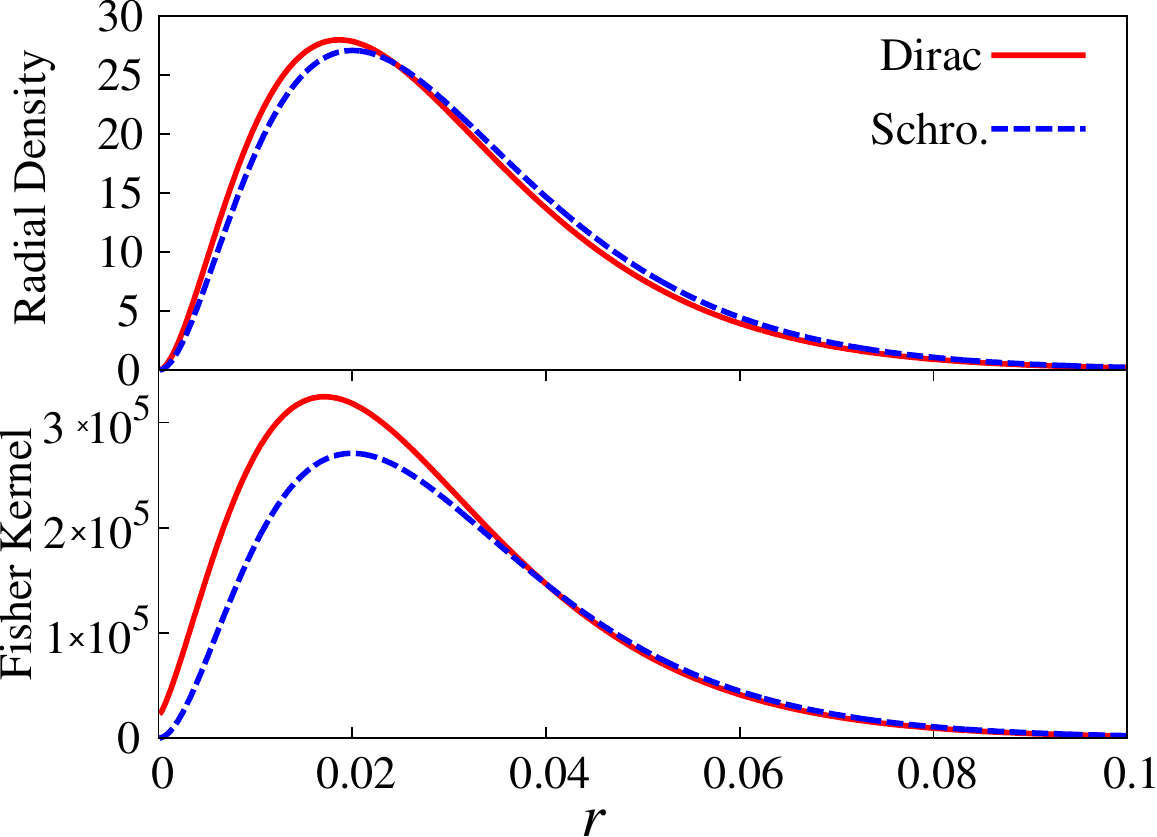}
\label{fig:state_gs}
}
\subfigure{
\includegraphics[scale=0.625]{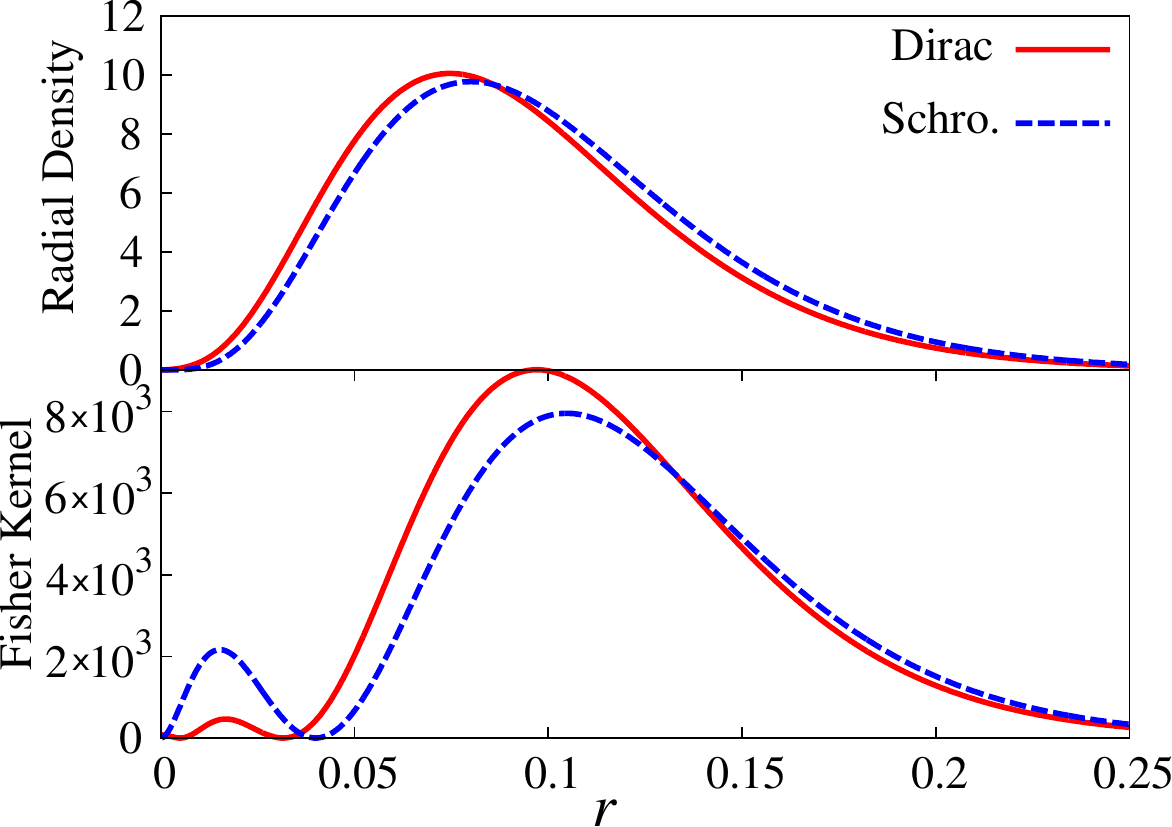}
\label{fig:state_circs}
}
\caption{(Color online) Radial density, $D^i(r)$, and radial Fisher information kernel, $I^i_{kernel}(r)$, in the Dirac ($i=D$) and Schr\"odinger ($i=S$) settings for the ground state (left) and the circular state $n=5$ (right) with nuclear charge $Z=50$. Atomic units have been used.}
\label{fig:dens_Ikernel_circ}
\end{figure}

For illustrative purposes we show in Fig. \ref{fig:dens_Ikernel_circ} the Dirac and Schr\"odinger radial distributions and the Fisher information kernels of the ground state and the circular state with $n = 5$ of the hydrogenic system with $Z = 50$. Therein it is observed that (i) in the two states the Dirac (solid red) radial density is always above the Schr\"odinger (dashed blue) curve when $r$ is less than the radial expectation value (centroid), and below otherwise, and (ii) the behavior of the Fisher kernel in the excited state is different from that of the ground state: the Dirac values are smaller than the Schr\"odinger ones not only when $r$ is bigger than the radial expectation value, but also in the neighbourhood of the nucleus. 

We have observed that the latter effect (to be called gradient reduction effect heretoforth) is present in all bound states other than the ground state, although in excited non-circular states this effect is hidden by the nodal disapearance or minima-raising effect. In circular states other than the ground state, this effect gives rise to the small negativity of the Fisher-Shannon ratio, as we can also observe in the next Fig. \ref{fig:zetaFS_mj} discussed in part B of this section.

Finally let us emphasize that while the LMC ratio quantifies the charge contraction towards the nucleus (mainly by means of its disequilibrium ingredient), the  Fisher-Shannon ratio quantifies the combined balance of this charge concentration, the gradient reduction in the regions near and far from the origin, and the minima raising or nodal disapearance of the charge distribution. This balance is very delicate, so that the latter ratio is positive in all ground-state systems and in all excited states of heavy hydrogenic states. However, the Fisher-Shannon ratio is negative for all excited states of hydrogenic systems with nuclear charge less than a critical state-dependent value; in these cases the relativistic minima-raising and gradient reduction joint effects are greater than the charge-contraction effect.

All in all, the Fisher-Shannon ratio quantifies (a) the charge contraction towards the nucleus in the ground state, (b) the charge contraction together with the gradient reduction effect for circular states other than the ground state, and (c) the combined effect due to the charge contraction, the gradient reduction and the minima raising for the remaining excited states.

\subsection{Dependence on quantum numbers}

\begin{figure}[ht]
\centering
\subfigure{
\includegraphics[scale=0.625]{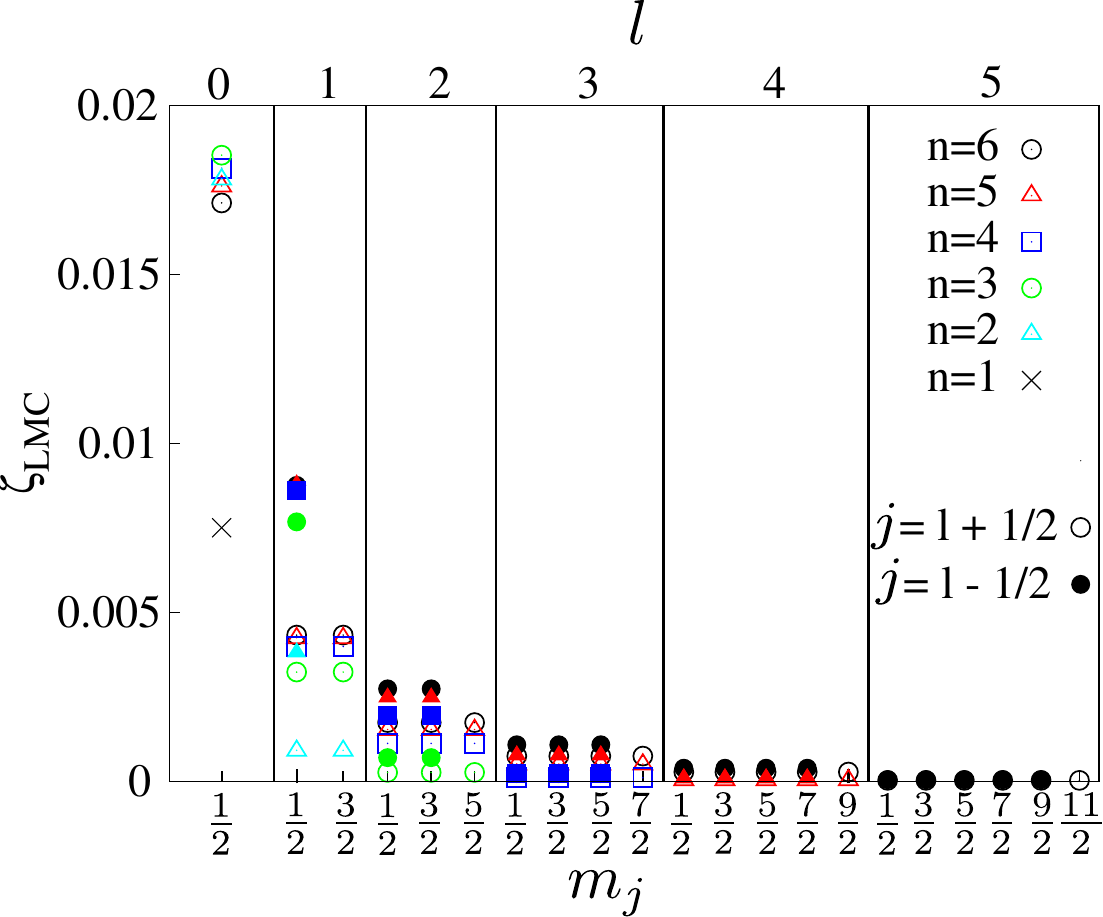}
\label{fig:zetaLMC19_mj}
}
\subfigure{
\includegraphics[scale=0.625]{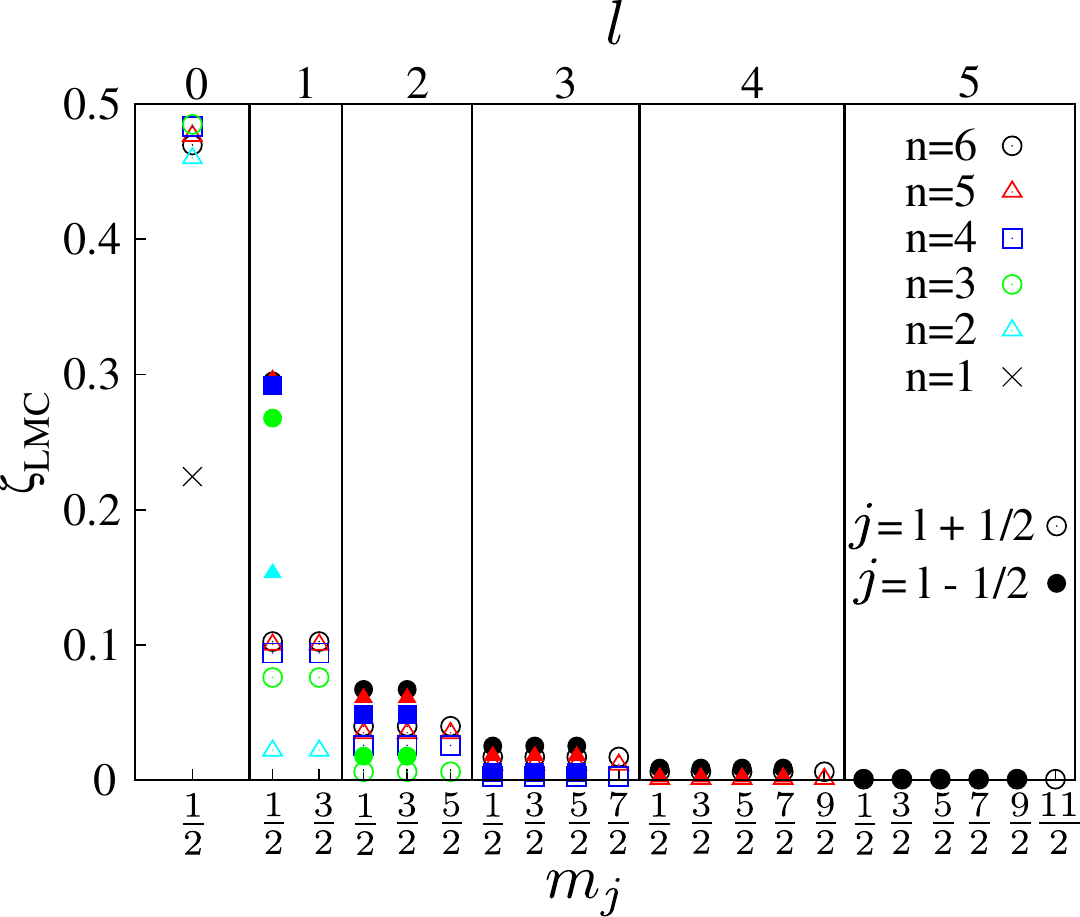}
\label{fig:zetaLMC90_mj}
}
\caption{(Color online) Dependence of the LMC complexity on $m_j$ for $Z=19$ (left) and $Z=90$ (right).}
\label{fig:zetaLMC_mj}
\end{figure}

\begin{figure}[ht]
\centering
\subfigure{
\includegraphics[scale=0.625]{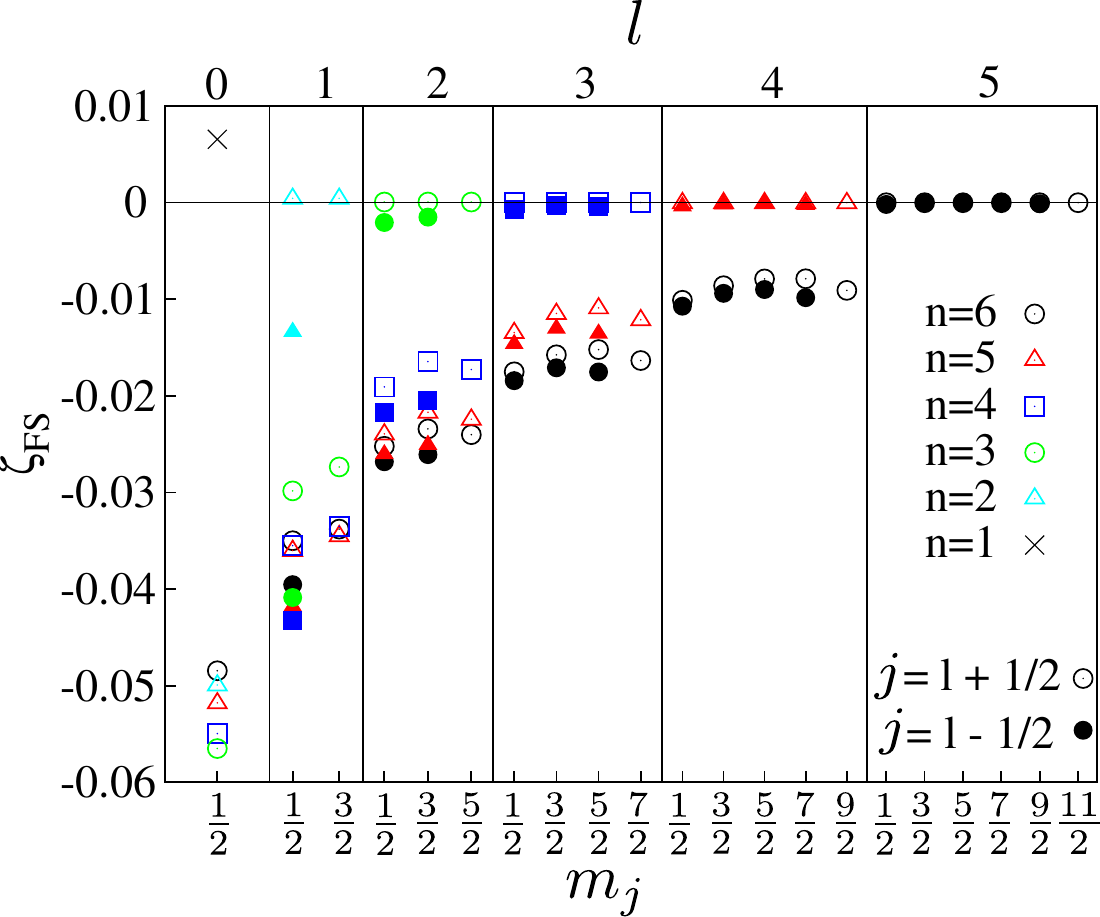}
\label{fig:zetaFS19_mj}
}
\subfigure{
\includegraphics[scale=0.625]{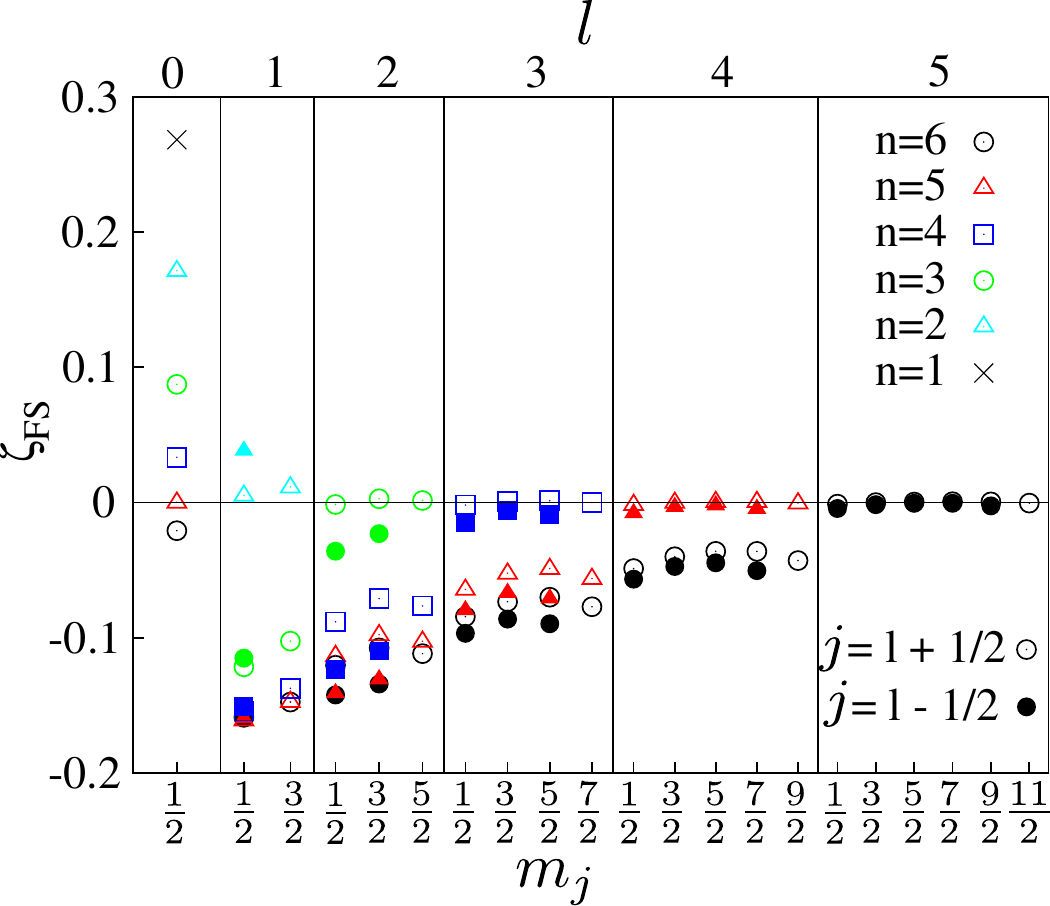}
\label{fig:zetaFS90_mj}
}
\caption{(Color online) Dependence of the Fisher-Shannon complexity on $m_j$ for $Z=19$ (left) and $Z=90$ (right).}
\label{fig:zetaFS_mj}
\end{figure}

First, the quantification of the Dirac effects for all excited states with $n\le 6$ in hydrogenic sytems with $Z=19$ and $Z=90$ is examined by means of the LMC (see Figs. \ref{fig:zetaLMC_mj}) and Fisher-Shannon (see Figs. \ref{fig:zetaFS_mj}) complexity ratios. In Fig. \ref{fig:zetaLMC_mj}-left for $Z = 19$ and Fig. \ref{fig:zetaLMC_mj}-right for $Z = 90$, the LMC ratio shows a common general structure. For given quantum numbers ($n,l$) the ratio has higher values for states with $j = l-\frac{1}{2}$ than for states with $j = l+\frac{1}{2}$. Moreover, it does not depend on the magnetic quantum number $m_j$,  what can be theoretically understood from Eqs. \eqref{eq:def_LMC}, \eqref{eq:def_D-S}, \eqref{eq:dirac_dens} and \eqref{eq:Sch_dens} which allow us to separate the LMC complexity as a product of a radial complexity (associated to the radial density) and an angular complexity (associated to the angular density, which is the same in both Dirac and Schr\"odinger frameworks); then, the LMC ratio has no 
dependence of any angular property. In addition, the LMC ratio (i) decreases when the orbital quantum number $l$ is increasing for fixed $n$ and (ii) increases with the principal quatum number $n$ for fixed values of ($j,l$). As already pointed out in part \ref{section3A}, for large values of $Z$, the bigger the nuclear charge is, the higher the ratio due to the common electronic charge contraction.

The Fisher-Shannon ratio presents a different behavior with respect to the quantum numbers than that of the LMC one, as we show in Fig. \ref{fig:zetaFS_mj}-left for $Z = 19$ and in Fig. \ref{fig:zetaFS_mj}-right for $Z = 90$.  Indeed, it has negative values except in a few s and p states. Moreover, although the relativistic effects are stronger in the system with nuclear charge Z = 90,  the qualitative dependence of the ratio on the quantum numbers is similar in the two systems: it has higher values for states with $j=l+\frac{1}{2}$ than for states with $j=l-\frac{1}{2}$ when ($n,l$) are fixed. For states with $l < n-1$ the ratio severely decreases because of the minima raising, as previously discussed. Moreover, for penetrating states (mainly, states $s$) the charge contraction effect counterbalances the minima-raising effect and makes the ratio to become positive. Besides, the ratio hardly depends on the magnetic quantum number $m_j$ because the Fisher-Shannon complexity, opposite to the LMC quantity, 
cannot be separated into radial and angular parts.

The gradient reduction effect is increasingly higher for states with $j=l-\frac{1}{2}$ than for states with $j=l+\frac{1}{2}$ when the nuclear charge is increasing. For excited states with $l<n-1$, the minima-raising effect (which grows with $Z$) decreases the ratio. For large values of $Z$, the charge-contraction effect is so powerful that makes negligible the gradient reduction and minima-raising effects, producing a global positive Fisher-Shannon ratio.

\begin{figure}[ht]
\centering
\subfigure{
\includegraphics[scale=0.625]{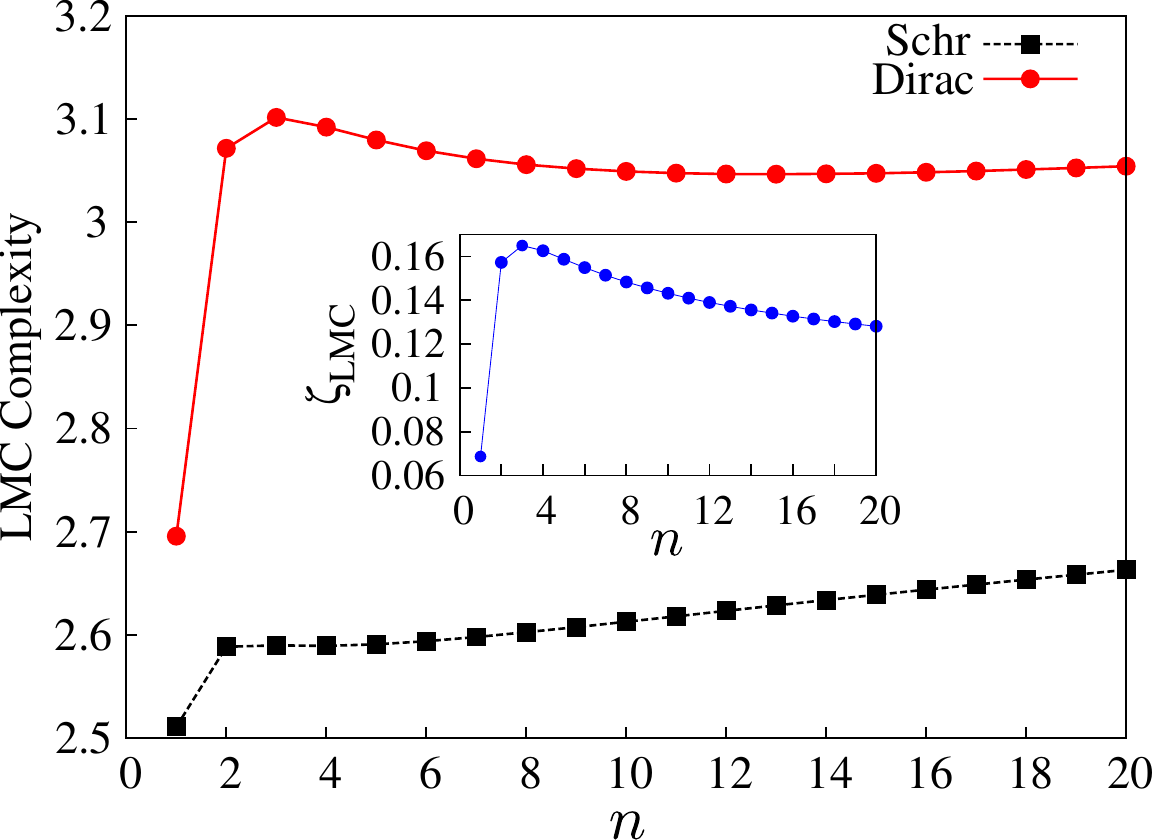} 
\label{fig:LMC(n)}
}
\subfigure{
\includegraphics[scale=0.625]{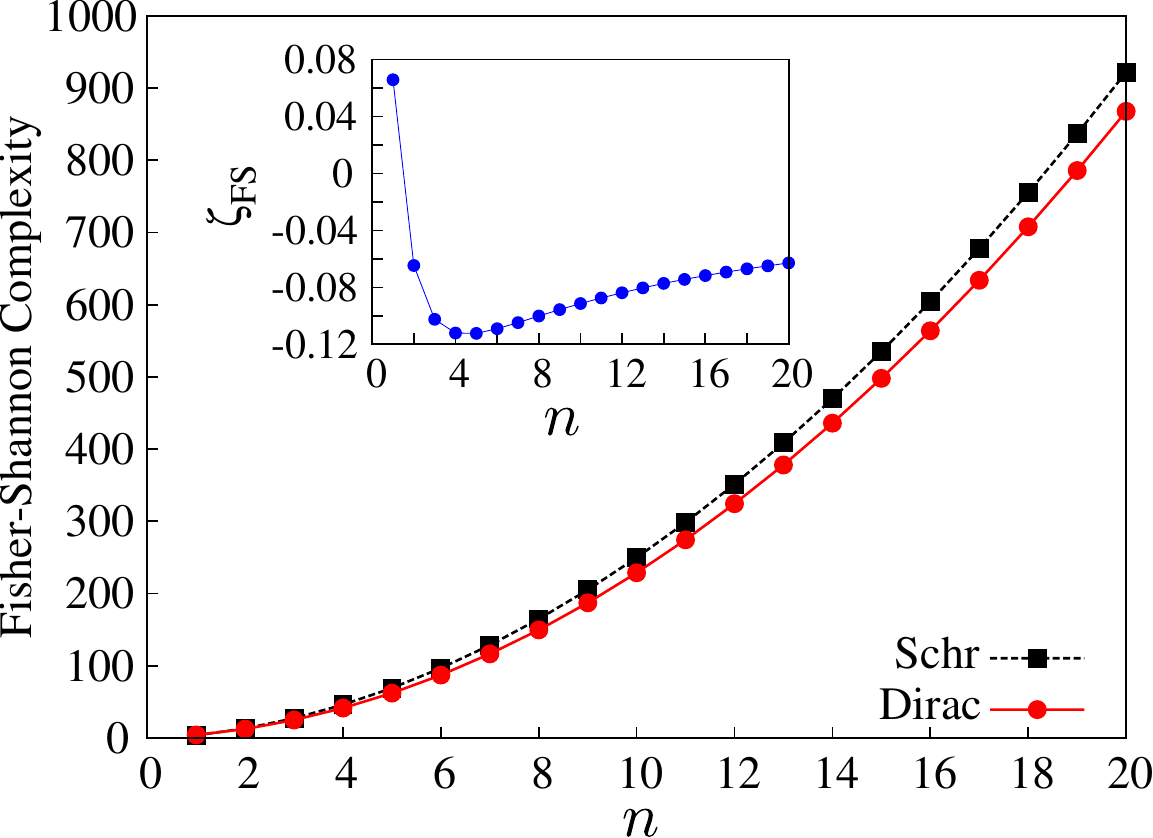} 
\label{fig:FS(n)}
}
\caption{(Color online) LMC (left) and Fisher-Shannon (right) complexities for excited s-states in $n$ with $Z=55$.}
\label{fig:C(n)}
\end{figure}

Second, we have done a similar analysis for states $n$s, which all have $j = l+\frac{1}{2}$, as shown in Figs. \ref{fig:C(n)} for the hydrogenic system with $Z = 55$. Contrary to the other excited states wherein the LMC (Fisher-Shannon) ratio decreases (increases) asymptotically to a constant value, the LMC ratio grows up to a maximum at $n = 3$, and then slowly decreases towards a constant asymptotic value as shown in Fig. \ref{fig:C(n)}-left. On the other hand, the Fisher-Shannon ratio (see Fig. \ref{fig:C(n)}-right) shows an opposite behavior as a function of $n$; that is, initially it decreases down to a minimun at $n=4$ and then it slowly increases towards a constant asymptotic value. For large values of $Z$, both LMC and Fisher-Shannon ratios of $n$s-states behave like in the other states. Notice, in addition, that LMC and Fisher-Shannon complexities of states $n$s have different behavior: while the LMC remains practically constant, the Fisher-Shannon increases monotonically. The latter is because the 
charge density oscillates more and more when the principal quantum number $n$ is increasing, what makes the Fisher-information factor of the Fisher-Shannon complexity to grow in a monotonic manner. 

\section{Complexity dependence on energy and Dirac quantum number}
\label{section4}

In this section we study the dependence of the LMC and Fisher-Shannon complexities on both the binding energy $B$ and the Dirac or relativistic quantum number $k$ of various excited states of the hydrogenic system with nuclear charge $Z = 90$, as well as we discuss the associated Fisher-Shannon ($I-J$) and disequilibrium-Shannon ($D-e^S$) information planes.

\subsection{Dependence on energy}
\label{section4A}

\begin{figure}[ht]
\centering
\subfigure{
\includegraphics[scale=0.625]{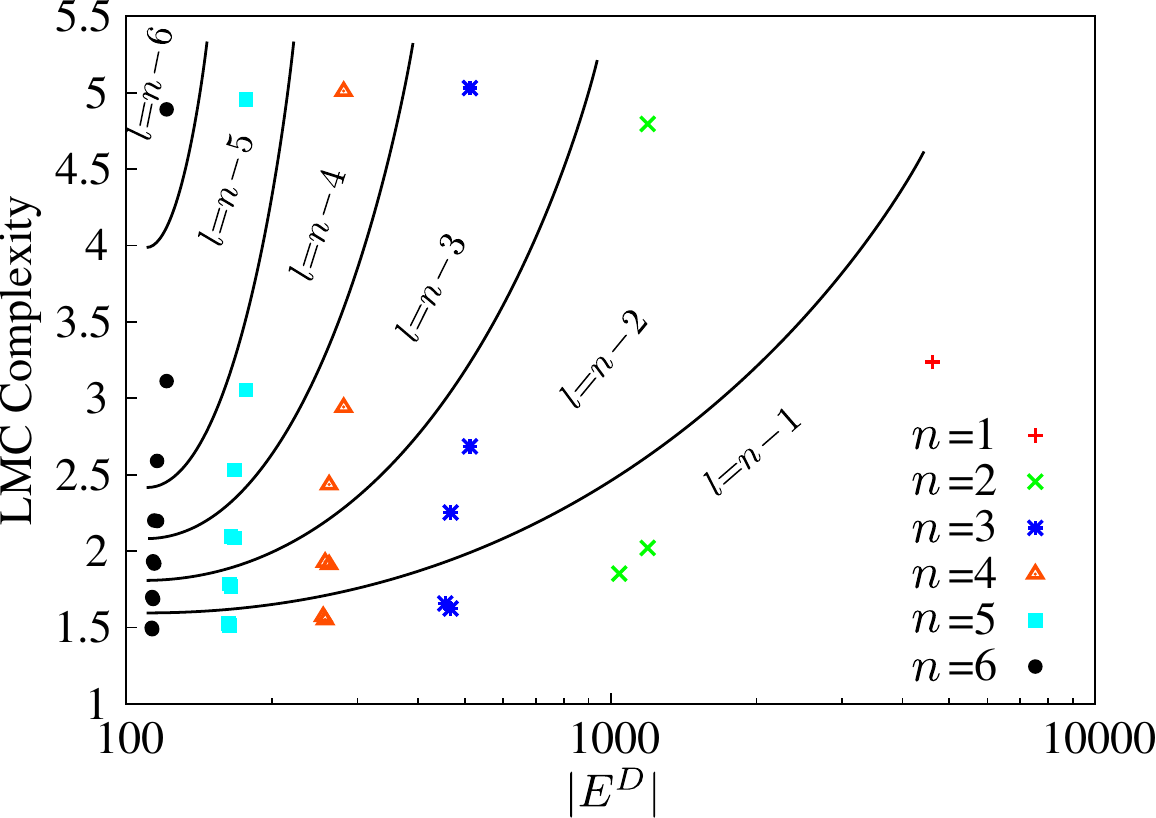} 
\label{fig:LMC(E)}
}
\subfigure{
\includegraphics[scale=0.625]{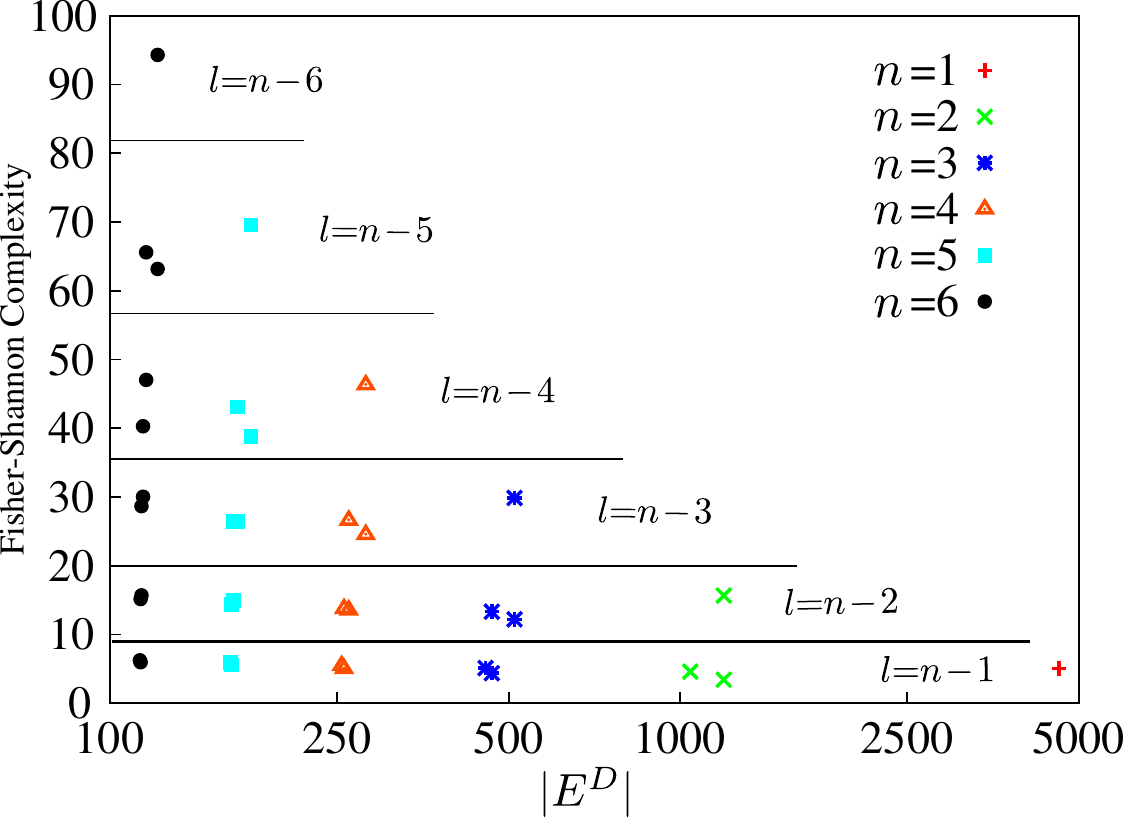} 
\label{fig:FS(E)}
}
\caption{(Color online) LMC (left) and Fisher-Shannon (right) complexities for some excited states in $n$ and $l$ with $Z=90$ and $m_j=j$ as a function of the energy (a.u.).}
\label{fig:C(E)}
\end{figure}

In Fig. \ref{fig:C(E)} we show the values of LMC (Fig. \ref{fig:C(E)}-left) and Fisher-Shannon (Fig. \ref{fig:C(E)}-right) complexities for all excited states $(n\le6, m_j = j)$ of the hydrogenic system with nuclear charge $Z = 90$. Therein we observe that when the energy is increasing, the LMC complexity (a) decreases for states with the same quantum number $j$, and (b) increases parabolically for states with $l=n-i$ and fixed $i$ $(i=1,...,n)$. Moreover, the LMC complexity of the states $n$s have significantly bigger values, mainly because of the relativistic sensitivity of the disequilibrium ingredient previously discussed.

Furthermore, the behavior of the Fisher-Shannon complexity as a function of the energy is similar to the LMC complexity for states with the same $j$, but it is slightly varying within a narrow interval for states with $l=n-i$ and fixed $i$ $(i=1,...,n)$.

\subsection{Dependence on the relativistic quantum number $k$}
\label{section4B}

\begin{figure}[ht]
\centering
\subfigure{
\includegraphics[scale=0.625]{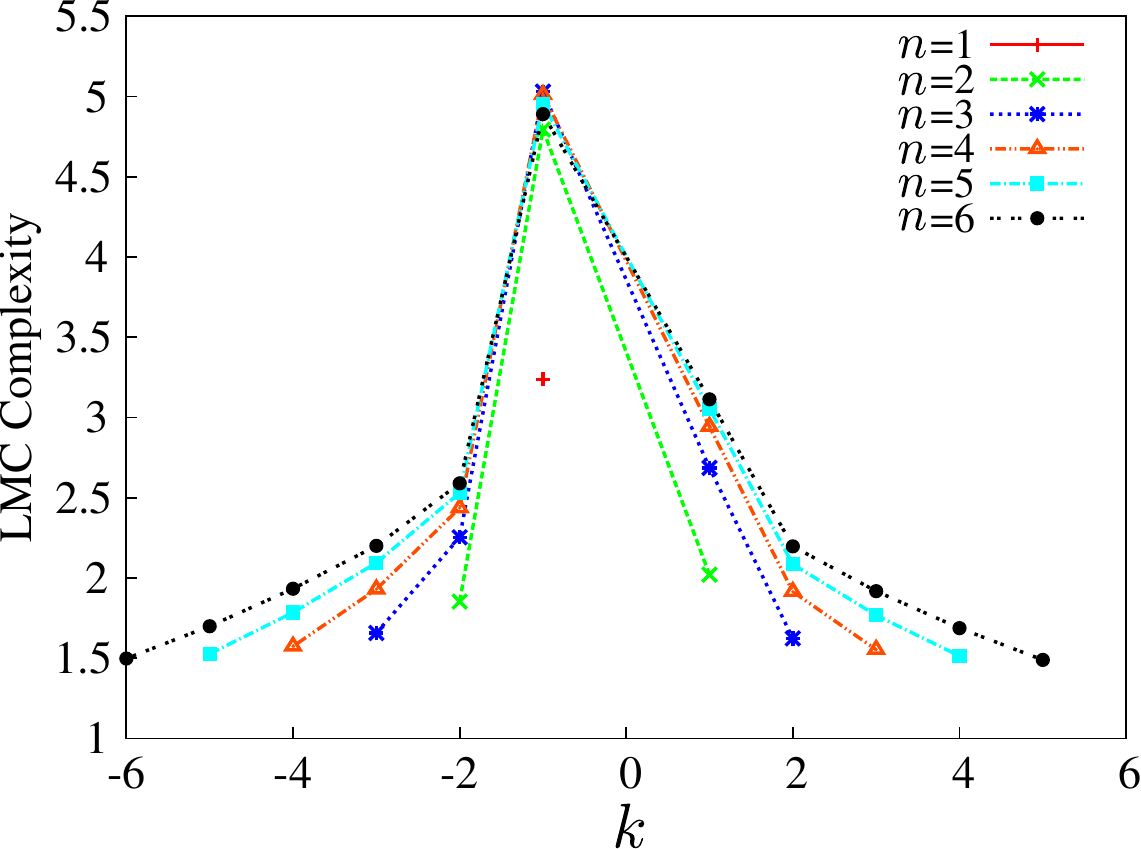} 
\label{fig:LMC(k)}
}
\subfigure{
\includegraphics[scale=0.625]{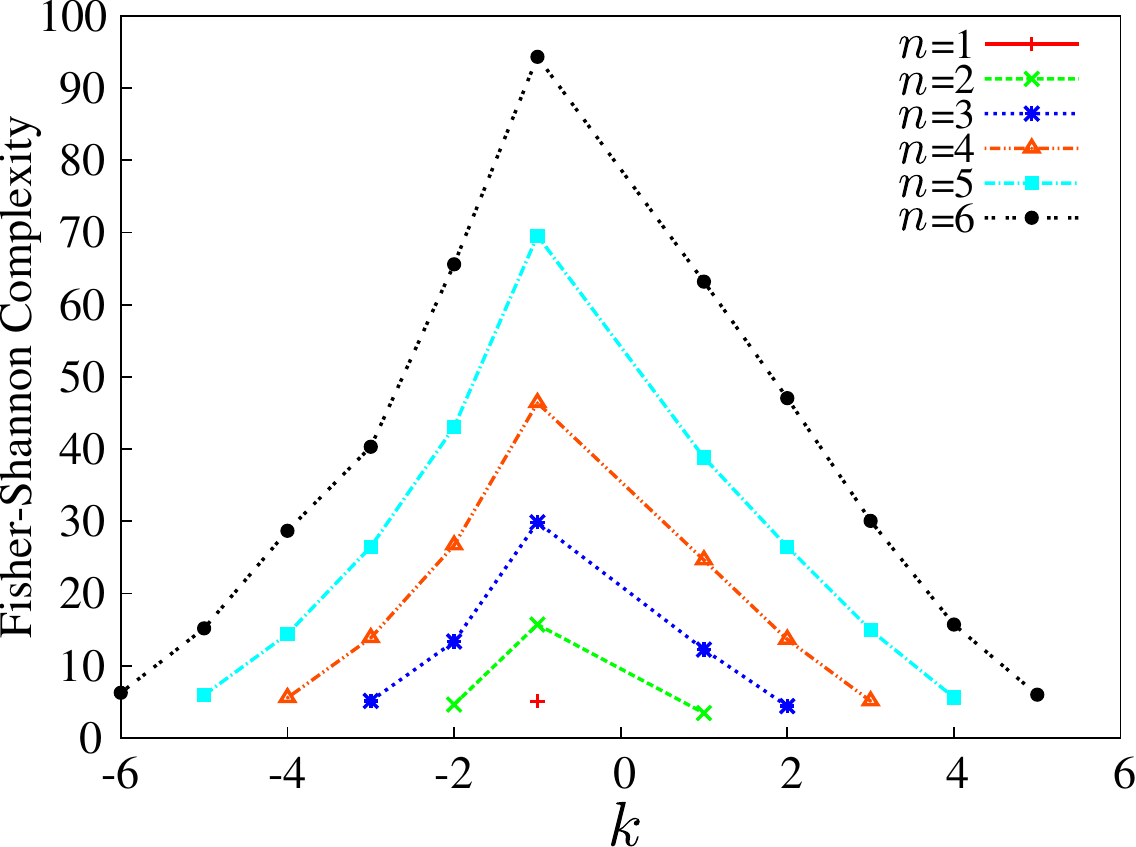} 
\label{fig:FS(k)}
}
\caption{(Color online) LMC (left) and Fisher-Shannon (right) complexities for some excited states in $n$ and $k$ with $Z=90$ and $m_j=j$ as a function of the relativistic quantum number $k$.}
\label{fig:C(k)}
\end{figure}

In Fig. \ref{fig:C(k)} we show the dependence of the LMC (left) and Fisher-Shannon (right) complexity measures on the relativistic quantum number $k$ for the ground state and all excited states ($n\le6,l,j,m_j=j$) of the hydrogenic system with nuclear charge $Z = 90$. 

We observe that the LMC complexity (a) has a global maximum for states $n$s (i.e., $k=-1$), and (b) presents a quasi-symmetric decreasing behavior around the line with $k=-1$ (i.e., for states $n$s). Moreover, the Fisher-Shannon complexity has not a global maximum at $n$s states but  it shows up a monotonically decreasing behavior for the $l$-manifold states with a given principal quantum number $n$, mainly because of the decreased number ($n-l$) of maxima of the density.

\subsection{Information planes}

\begin{figure}[ht]
\centering
\subfigure{
\includegraphics[scale=0.62]{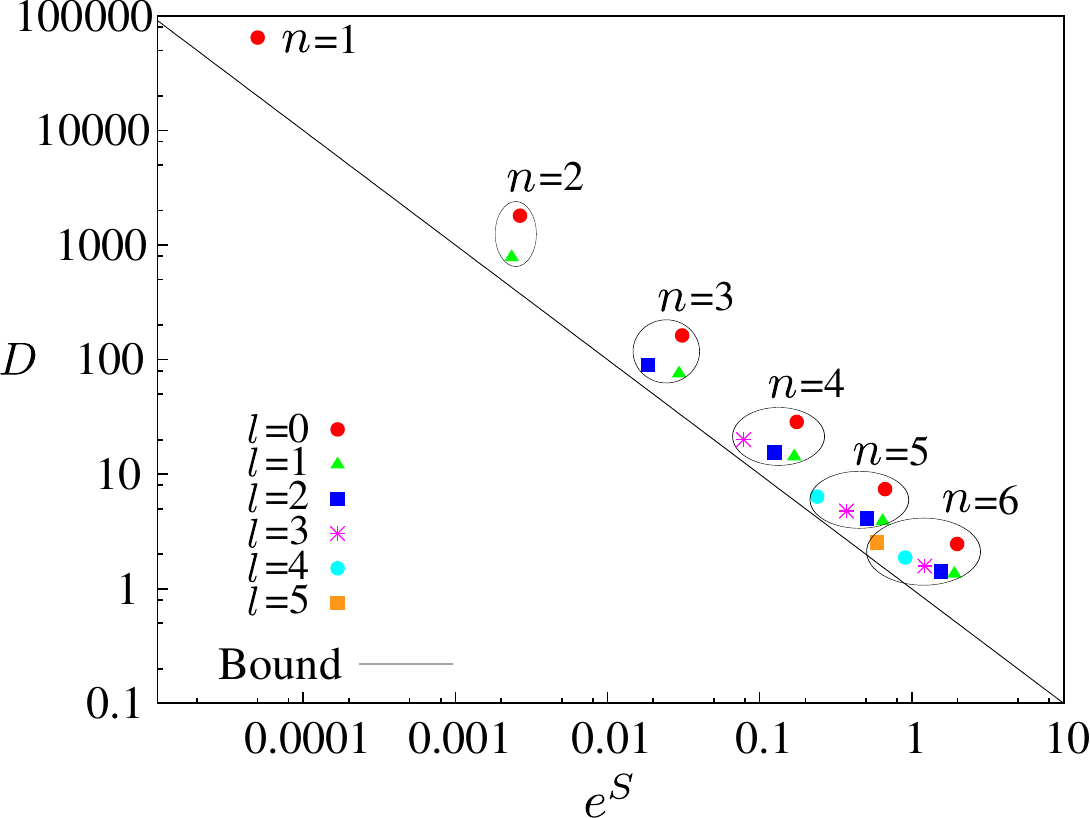} 
\label{fig:PlanosInfoD-S}
}
\subfigure{
\includegraphics[scale=0.62]{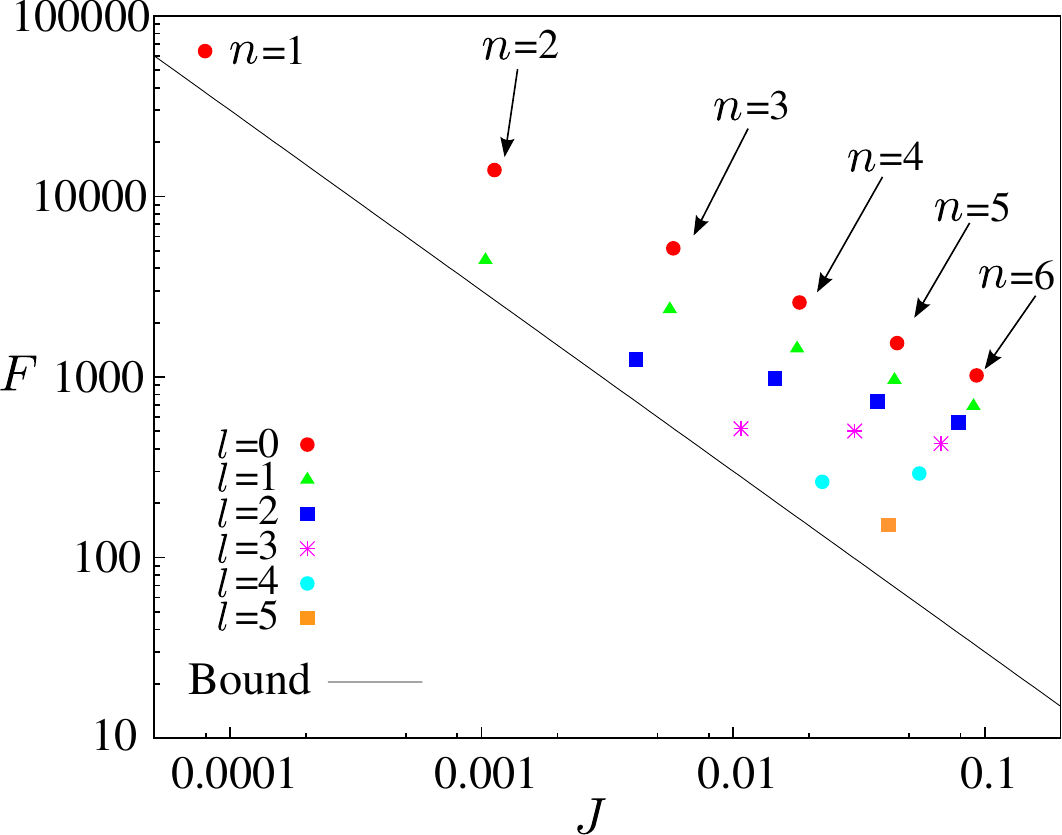} 
\label{fig:PlanosInfoF-S}
}
\caption{(Color online) LMC or Disequilibrium-Shannon (left) and Fisher-Shannon (right) information planes of hydrogenic states with $n\le6$, $m_j=j=l+\frac{1}{2}$ and $Z=90$. Atomic units have been used.}
\label{fig:InfoPlanes}
\end{figure}

Finally, it is interesting to remark that the previous behavior can be studied by means of the associated relativistic information-theoretic planes. In Fig. \ref{fig:InfoPlanes} we show the Disequilibrium-Shannon (left) and Fisher-Shannon (right) information planes which include the ground state and all excited states ($n\le6,l,j=l+\frac{1}{2},mj=j$) of the hydrogenic system with nuclear charge $Z=90$. Notice that the scale in both axes is logarithmic.

First of all, we observe that in both cases all the complexity values lie down the allowed region; that is, they are in the right side of the rigorous border (see continuous line in the two graphs) defined by the known analytic LMC and Fisher-Shannon lower bounds \cite{lopezrosa:pa09, stam:ic59, guerrero:pra11}: $C_{\text{LMC}} \left[ \rho \right] \geq 1$ and $C_{\text{FS}} \left[ \rho \right] \geq 3$. Moreover while the LMC values remain closer to the borderline, the Fisher-Shannon ones move away from this bound when the principal quantum number is increasing. This is a clear indication that the Fisher-Shannon values of a given state (a) are higher than the corresponding LMC ones and (b) this enhancement is  greater when the principal quantum number is increasing, mainly because the gradient content (so, the Fisher-information ingredient) raises in a faster manner than the disequilibrium.

\section*{Conclusions} 

This work extends the information-theoretic study of the hydrogenic systems recently done in the Schr\"odinger \cite{dehesa:epjd09} and relativistic Klein-Gordon \cite{manzano:njp10,manzano:epl10} and Dirac \cite{katriel:jcam10} frameworks. Indeed, previous efforts have analyzed not only the single and composite information-theoretic measures of both ground and excited states in the Schr\"odinger \cite{dehesa:ijqc10, dehesa:epjd09}] and Klein-Gordon \cite{manzano:njp10,manzano:epl10} settings, but also the single entropic measures of the ground state in the Dirac setting \cite{katriel:jcam10}. Here we have studied the LMC and Fisher-Shannon complexities of both ground and excited states of these systems by means of the Dirac relativistic wavefunctions. First we have shown the enhancement of these composite measures when the nuclear charge is increasing and we have compared these values with the corresponding non-relativistic (Schr\"odinger) ones, what has allowed us to (i) illustrate that these complexity 
measures are good indicators of the Dirac relativistic effects, and (ii) to quantify the three primary dinamical Dirac effects (electronic charge contraction, minima-raising and gradient reduction) by means of a Schr\"odinger-Dirac ratio. We have observed that while the LMC ratio is always positive and it has an increasing behavior as a function of $Z$ (mainly because its disequilibrium ingredient enhances when $Z$ is increasing), the Fisher-Shannon ratio can reach negative values for the excited states although finally enhances when $Z$ is increasing. Moreover, the global enhancement phenomenon of the two complexities is mainly due to the electronic charge contraction, and the Fisher-Shannon negativity in the excited states is associated to the raising of the non-relativistic minima. The latter phenomenon is mainly due to the Fisher-information ingredient of the Fisher-Shannon complexity, because it is the only factor which is very sensitive to the fact that the Dirac relativistic radial density cannot 
vanish except at the origin and infinity, keeping in mind that it is a gradient functional of the density. The (largely ignored) gradient reduction effect is present in all excited states although it is, at times, hidden by the minima-raising effect.

Furthermore, we have shown in a large-$Z$ hydrogenic system the dependence of the two previous statistical complexities as a function of the following parameters of the Dirac states: the energy, the principal quantum number ($n$) and the relativistic quantum number ($k$). We have observed that for the $l$-manifold states of a given quantum number $n$,  the LMC complexity parabolically enhances  and the Fisher-Shannon complexity varies within the same interval when energy is increasing; this is mainly because of the delicate balance of the charge contraction and the minima raising effects. Besides, beyond the ground state, we have observed that for $j=m_j$ the behavior of the two complexity measures of the $l$-manifold states in terms of the relativistic quantum number $k$ is quasi-symmetric around the line with $k=-1$ (i.e., states s).

\section*{Acknowledgments}
This work was partially supported by the Excellence Projects FQM-2445 and FQM-4643 of the Junta de Andaluc\'ia (Spain, EU), and the grant FIS2011-24540 of the Ministerio de Innovaci\'on y Ciencia. We belong to the
Andalusian research group FQM-207 (J.S.D), FQM-020 (P.A.B) and FQM-239 (S.L.R).


\end{document}